\documentstyle[12pt]{article}
\def\del{\partial}
  \makeatletter
        
        \@addtoreset{equation}{section}
        \def\section{\@startsection {section}{1}{\z@}{3.5ex plus -1ex minus
        -.2ex}{2.3ex plus .2ex}{\large\bf}}
  \makeatother
\begin{document}
\setlength{\baselineskip}{22pt}
\rightline{YAMAGATA-HEP-93-13}
\rightline{August 1993}
\vspace{1.5truecm}
        \renewcommand{\thefootnote}{\fnsymbol{footnote}}
\centerline{\Large\bf Gaugeon Formalism with BRST Symmetry\footnote[2]{%
    Preliminary result was reported at Niigata Autumn School on
    Particle Physics held on October 23--25, 1992, in Kusatsu.}}
\vspace{2truecm}
\centerline{Minoru KOSEKI, Masaaki SATO$^*$ and Ryusuke ENDO$^{**}$}
\bigskip
\centerline{{\it Graduate School of Science and Technology }}
\centerline{{\it Niigata University, Niigata 950-21, Japan}}
\centerline{{\it $^*$Hitachi Tohoku Software Co.,
                    Sendai 980, Japan}}
\centerline{{\it $^{**}$Department of Physics, Yamagata University,
                         Yamagata 990, Japan}}
\vspace{2truecm}
\centerline{\bf Abstract}
\bigskip
We provide a BRST symmetric version of Yokoyama's Type I gaugeon formalism
for quantum electrodynamics; the similar theory by Izawa can be considered
as a BRST symmetrized Type II theory. With the help of the BRST symmetry,
Yokoyama's physical subsidiary conditions are replaced by the
Kugo-Ojima type condition.
As a result, the formalism becomes applicable even in the background
gravitational field.
We show how the Hilbert spaces of standard formalism in various gauges
are embedded in the single Hilbert space of the present formalism.
We also give a path integral derivation of the Lagrangian.

\thispagestyle{empty}
\newpage
\renewcommand{\thefootnote}{\arabic{footnote}}
\section{Introduction}
In the standard formalism of canonically quantized gauge theories
\cite{N,KO}
we cannot consider the gauge transformation freely.
There exists no gauge freedom in the quantum theory, since the
theory is defined only after the gauge fixing. Namely, a Hilbert space
defined in a particular gauge is quite different from those in other gauges.
Thus, if we want to realize the quantum gauge freedom, we need a wider
Hilbert space.

Yokoyama's gaugeon formalism \cite{OEM}$-$\cite{YM2} provides a wider
framework in which we can consider the quantum gauge transformation
among a family of Lorentz covariant linear gauges. In this formalism
a set of extra fields, so called gaugeon field, is introduced as the quantum
gauge freedom. This theory was first proposed for the quantum
electrodynamics \cite{OEM}$-$\cite{txt}
to resolve the problem of gauge parameter renormalization \cite{HY}.
It was also applied later to the Yang-Mills theory \cite{YM1,YM2}.
Thanks to the quantum gauge freedom of this formalism, the gauge parameter
independence of the physical $S$-matrix becomes manifest \cite{Smatrix}.
It has also been shown, with the help of certain conjecture,
that the wave-function renormalization constant is gauge independent
in this formalism \cite{Zfactor}.

The extra gaugeon modes should be removed from the physical
Hilbert space since the quantum gauge freedom is unphysical mode.
In fact the gaugeon exhibits dipole character and yields negative normed
states. To remove these modes Yokoyama imposed the Gupta-Bleuler type
subsidiary condition \cite{OEM}.
However, this type of condition does not work well if interaction
is present for the gaugeon field. Especially, we cannot use Yokoyama's
subsidiary condition in the background gravitational field.

In the present paper we improve the subsidiary conditions of Yokoyama's
formalism for the quantum electrodynamics. For this purpose we introduce
extra Faddeev-Popov [FP] ghosts for the gaugeon field and provide
a fully BRST symmetric Lagrangian.
By the help of BRST charge the subsidiary condition is
simply replaced by the Kugo-Ojima type condition \cite{KO}, which is known
to work well even in the interacting case.

In connection with this program, we should refer to the work by
Izawa \cite{Izawa}. After completed our work, we became aware of his
paper in which he also provided a BRST symmetric gaugeon formalism and
the Kugo-Ojima type subsidiary condition. The relation of
our theory to his is also discussed in this paper.

The paper is organized as the following. In \S2 we briefly review
Yokoyama's gaugeon formalism for quantum electrodynamics. In \S3 a
BRST symmetrized version of this formalism is proposed. And its relation
to Izawa's theory is discussed. In \S4 we see how the Hilbert space of the
standard formalism is embedded in the wider space of the BRST symmetric
gaugeon formalism.
To confirm the necessity of FP ghosts for the gaugeon field,
we re-derive our Lagrangian and Izawa's Lagrangian using path
integral in \S5. Section 6 is devoted to summary.

\section{Yokoyama's gaugeon formalism}

Yokoyama's Lagrangian for electromagnetic field $A_\mu$ interacting with
charged matter field $\psi$ is given by
\begin{eqnarray}
  {\cal L}_{\rm I}
          &=&
           - {1 \over 4}F_{\mu \nu }F^{\mu \nu }
           + \partial_\mu B A^\mu
           + \partial_\mu Y_* \partial^\mu Y
           + {\varepsilon \over 2} \left( Y_* + \alpha B \right)^2
                                                 \nonumber \\
       & & - i \partial_\mu c_* \partial^\mu c
           + {\cal L}_{\rm matt}(\psi, A_\mu) ,
                                                 \label{LI}
\end{eqnarray}
where $F_{\mu\nu}=\partial_\mu A_\nu-\partial_\nu A_\mu$, $B$ is the
$B$-field of Nakanishi-Lautrup \cite{N}, $c$ and $c_*$ are usual FP ghosts,
$Y$ and $Y_*$ are scalar fields called the gaugeon fields and its associated
field respectively\footnote{
               We use field notation different from Yokoyama's;
               the fields $B$, $Y$ and $Y_*$ here are denoted by $B_1$,
               $B$ and $B_2$, respectively, in the original paper.\cite{OEM}
                           }
and ${\cal L}_{\rm matt}$ is the Lagrangian for the
matter field $\psi$ minimally coupled to $A_\mu$.
We have introduced here the FP ghosts $c$ and $c_*$ which did not appear
in the original paper \cite{OEM}. Yokoyama introduced them later in the
application of his formalism to the Yang-Mills theory \cite{YM1}.
Namely, we start from the Abelian limit of Yokoyama's Lagrangian for the
Yang-Mills field \cite{YM1,YM2}
In (\ref{LI}), $\varepsilon$ denotes a sign factor ($\varepsilon
=\pm 1$) and $\alpha$ is a numerical gauge parameter.
The gauge parameter of the standard formalism, which we denote by $a$ in the
present paper, can be idetified with $a=\varepsilon\alpha^2$. For example,
Yokoyama's Lagrangian gives the photon propagator as
\begin{equation}
           {{g_{\mu\nu}}\over k^2}
                + (a - 1){k_\mu k_\nu \over (k^2)^2}
                \quad\quad \mbox{with~~$a=\varepsilon \alpha^2$.}
                                  \label{propagator}
\end{equation}
In particular, $\alpha=0$ corresponds to Landau gauge ($a=0$) and $\alpha=1$
with $\varepsilon=+1$ gives Feynman gauge ($a=1$). Note that the gaugeon
sector in (\ref{LI}) decouples from the rest if $\alpha=0$.
Then the remaining sector has the same form with the Lagrangian of the
standard formalism in Landau gauge.

The Lagrangian permits the q-number gauge transformation which
enables us to vary gauge parameter. The transformation is defined by
\begin{eqnarray}
   & &\hat A_\mu = A_\mu + \tau \partial _\mu Y, \nonumber \\
   & &\hat \psi = e^{i\tau eY} \psi,             \nonumber \\
   & &\hat Y_* = Y_* - \tau B,                   \nonumber \\
   & &\hat B = B, \quad \hat Y = Y,             \nonumber \\
   & &\hat c = c, \quad \hat c_* = c_*,
                                                  \label{qgt1}
\end{eqnarray}
with $\tau$ being a parameter of the transformation. Under this transformation
the Lagrangian (\ref{LI}) is {\it form-invariant}, that is, it transforms as
\begin{equation}
             {\cal L}_{\rm I}(\phi^A; \alpha)
                =  {\cal L}_{\rm I}(\hat \phi^A; \hat \alpha),\label{forminvI}
\end{equation}
where ${\phi^A}$ stands for any of the fields and $\hat\alpha$ is defined by
\begin{equation}
          \hat \alpha = \alpha + \tau .   \label{hat_alpha}
\end{equation}
The form invariance (\ref{forminvI}) means that $\phi^A$ and
$\hat \phi^A$ satisfy
the same field equation except for the parameter $\alpha$ which should be
replaced by $\hat \alpha$ for the $\hat \phi^A$ field equation. In this sense,
we can shift the gauge parameter by this transformation.
For example, starting from any value of $\alpha$,
we can always take $\alpha=0$ gauge where the theory is equivalent to
the Landau-gauge standard formalism (plus free $Y$-$Y_*$ system).

As subsidiary conditions to confine the unphysical modes, Yokoyama \cite{YM1}
adopted
\begin{eqnarray}
    & &Q_{\rm B(KO)} \left| \rm phys \right\rangle = 0,
                                               \label{BRSphys} \\
    & &Y_*\!^{(+)}(x) \left| {\rm phys} \right\rangle=0,
                                               \label{Y*phys}
\end{eqnarray}
where $Q_{\rm B(KO)}$ is the usual BRST charge \cite{KO} and $Y_*\!^{(+)}(x)$
is the positive frequency part of $Y_*(x)$. The first condition is the usual
one of the standard formalism, which confine the unphysical photons by the
quartet mechanism \cite{KO}. The second removes the gaugeon modes. It is
essential in the second condition that the $Y_*$ field satisfies the
free field equation\footnote{In
           the case of Yang-Mills field, $Y_*$ dose not satisfy the free field
           equation. Instead of it, however, the combination $Y_* + \alpha B$
           satisfies the free field equation. And the condition (\ref{Y*phys})
           is replaced by
           $$
               (Y_* + \alpha B)^{(+)} \left|{\rm phys}\right\rangle=0
           $$
           in Yang-Mills case \cite{YM1,YM2}
           }
\begin{equation}
                     \Box Y_* = 0.
\end{equation}
If this does not hold, the positive frequency part $Y_*\!^{(+)}$ becomes
ambiguous and the condition (\ref{Y*phys}) contradicts with time evolution
in general. For example, we cannot use the condition (\ref{Y*phys}) in
the background curved space-time.

There is another type of Lagrangian for the gaugeon formalism. In Ref.
\cite{YK} Yokoyama and Kubo discussed more general Lagrangians that include
$B$, $Y$ and $Y_*$ fields. Their conclusion is that
there are only two types of the theory which admits the q-number gauge
transformation. The first type [Type I] is described by (\ref{LI}) and the
the second [Type II] by
\begin{equation}
   {\cal L}_{\rm II}= - {1 \over 4}F_{\mu\nu}F^{\mu\nu}
                    + \partial_\mu B A^\mu + {\alpha \over 2}B^2
                    + \partial_\mu Y_* \partial^\mu Y
                    + {1\over2} Y_* B + \cdots,
                                             \label{type2}
\end{equation}
where the dots denote the matter sector and possible FP ghost term.
The Lagrangian (\ref{type2}) is also form-invariant under the q-number
gauge transformation (\ref{qgt1}).
The $\alpha$ in ${\cal L}_{\rm II}$ directly corresponds to the gauge
parameter of the standard formalism: $a=\alpha$.
We can shift this parameter into arbitrary value by the q-number gauge
transformation.
This should be compared with Type I theory in which we cannot change the sign
of the gauge parameter $a=\varepsilon\alpha^2$.
In the Type II case, however, the gaugeon sector would not
decouple even if any value of the parameter $\alpha$ taken. Thus, the
equivalence of the theory to the standard formalism is not so trivial as
in the Type I theory.

\section{Gaugeon formalism with BRST symmetry}

We consider the Lagrangian
\begin{eqnarray}
 {\cal L}&=& - {1 \over4} F_{\mu \nu }F^{\mu \nu }
             + \partial_\mu B A^\mu
             + \partial _\mu Y_* \partial ^\mu Y
             + {\varepsilon\over2} ( Y_* + \alpha B )^2 \nonumber \\
         & & - i \partial_\mu c_* \partial^\mu c
             - i \partial_\mu K_* \partial^\mu K
             + {\cal L}_{\rm matt}(\psi,A_\mu),
                                            \label{LBRST}
\end{eqnarray}
where we have introduced the FP ghosts $K$ and $K_*$ for the gaugeon fields.
This Lagrangian is different from Yokoyama's (\ref{LI}) only in the term of
$K$ and $K_*$ fields.
Field equations which follow from (\ref{LBRST}) are
\begin{eqnarray}
      &  &\del^\mu F_{\mu\nu} + \del_\nu B + j_\nu =0, \label{eqA} \\
      &  &\del^\mu A_\mu = \varepsilon\alpha(Y_* + \alpha B), \label{divA} \\
      &  &\Box Y = \varepsilon(Y_* + \alpha B), \\
      &  &\Box Y_* = 0, \\
      &  &\Box c = \Box c_* = 0, \\
      &  &\Box K = \Box K_* = 0,   \label{eqK}
\end{eqnarray}
with $j_\mu$ being the conserved current defined by
$j_\mu=\del{\cal L}_{\rm matt}/\del A^\mu$.

The Lagrangian (\ref{LBRST}) is invariant under the following BRST
transformation,
\begin{eqnarray}
  & &\delta_{\rm B} A_\mu = \partial_\mu c, \nonumber \\
  & &\delta_{\rm B} \psi = ie c \psi,       \nonumber \\
  & &\delta_{\rm B} c_* = -iB,              \nonumber \\
  & &\delta_{\rm B} B = \delta_{\rm B}c=0,  \nonumber \\
  & &\delta_{\rm  B} Y = K,                 \nonumber \\
  & &\delta_{\rm B} K_* = -iY_*,            \nonumber \\
  & &\delta_{\rm B}Y_* =\delta_{\rm B}K=0,
                                            \label{BRST}
\end{eqnarray}
which obviously satisfies the nilpotency, $\delta_{\rm B}^{~2}=0$.
Because of the nilpotency, the BRST invariance is easily seen
if we rewrite the Lagrangian as
\begin{eqnarray}
  {\cal L} &=& - {1\over4} F_{\mu \nu }F^{\mu \nu }
               + {\cal L}_{\rm matt}                 \nonumber \\
           & & + i \delta_{\rm B}
         \left[
                 \partial_\mu c_* A^\mu
               + \partial_\mu K_* \partial^\mu Y
               + {\varepsilon\over2}(K_* + \alpha c_*)( Y_* + \alpha B)
         \right].
\end{eqnarray}
BRST charge corresponding to this transformation is expressed by
\begin{equation}
    Q_{\rm B} = \int ( c \mathop{\del_0}^\leftrightarrow B
                                  + K \mathop{\del_0}^\leftrightarrow Y_*
                                   )d^3x,             \label{QB}
\end{equation}
with
$      \stackrel{\leftrightarrow}{\del_0}
      = \del_0 - \stackrel{\leftarrow}{\del_0}$.
By the help of this charge we can define the physical subspace
${\cal V}_{\rm phys}$ as a space of states satisfying
\begin{equation}
     Q_{\rm B} \left|{\rm phys} \right\rangle = 0.  \label{QBphys}
\end{equation}
This subsidiary condition removes the gaugeon modes as well as the
unphysical photons from the physical subspace; $Y$ and $Y_*$ together with
$K$ and $K_*$ constitute a BRST quartet.

We consider the following q-number gauge transformation:
\begin{eqnarray}
  & &\hat A_\mu = A_\mu + \tau\del_\mu Y, \nonumber \\
  & &\hat \psi = e^{i\tau eY}\psi,        \nonumber \\
  & &\hat Y_* = Y_* - \tau B,             \nonumber \\
  & &\hat B = B, \quad \hat Y = Y,        \nonumber \\
  & &\hat c = c + \tau K,                 \nonumber \\
  & &\hat K_* = K_* - \tau c_*,           \nonumber \\
  & &\hat c_* = c_*, \quad \hat K = K,
                                          \label{qgt2}
\end{eqnarray}
Under this field transformation, the Lagrangian is form-invariant:
\begin{equation}
      {\cal L}(\phi^A,\alpha) = {\cal L}(\hat \phi^A,\hat\alpha),
\end{equation}
with $\hat \alpha=\alpha + \tau$. Thus, $\hat \phi^A$ also satisfies the field
equations (\ref{eqA})--(\ref{eqK}) with $\alpha$ replaced by $\hat \alpha$.

It should be noted that the q-number gauge transformation (\ref{qgt2})
commutes with the BRST transformation (\ref{BRST}). As a result, our BRST
charge (\ref{QB}) is invariant under the q-number transformation:
\begin{equation}
        \hat Q_{\rm B} = Q_{\rm B}.  \label{QB_hat}
\end{equation}
The physical subspace ${\cal V}_{\rm phys}$ is, therefore, invariant under
the q-number gauge transformation:
\begin{equation}
             \hat {\cal V}_{\rm phys} = {\cal V}_{\rm phys}.
\end{equation}
This situation does not occur in Yokoyama's partially BRST symmetric theory
\cite{YM1,YM2}.  Yokoyama's
condition (\ref{Y*phys}) is not invariant under the transformation
(\ref{qgt1}).

Before concluding this section, we refer to Izawa's theory \cite{Izawa}.
He has given a BRST symmetric Lagrangian which admits the q-number gauge
transformation. His Lagrangian is similar to ours, but slightly different.
Actually, it can be expressed in our notation as
\begin{equation}
   {\cal L}_{\rm Izawa} = {\cal L}_{\rm II}
                          - i \del_\mu K_* \del^\mu K. \label{LIzawa}
\end{equation}
Namely, Izawa's Lagrangian can be regarded as a BRST symmetric version
of the Type II gaugeon theory,\cite{YK} while ours is of Type I.
Especially, one cannot decouple the gaugeon sector from the Lagrangian by
choosing any value for the gauge parameter.
Thus, its equivalence to the standard formalism in the Landau gauge is not
so manifest as that of the Type I theory.\footnote{
                      The equivalence will be seen in the next section.
                                          }
\section{Gauge structure of Hilbert space}

As well as the BRST symmetry (\ref{BRST}), the Lagrangian (\ref{LBRST})
has several other symmetries. For example, we have the following BRST-like
conserved charges:
\begin{eqnarray}
    & &Q_{\rm B(KO)} = \int c \stackrel{\leftrightarrow}{\del_0} B d^3x,
				      \nonumber \\
    & &Q_{\rm B(Y)~} = \int K \stackrel{\leftrightarrow}{\del_0} Y_* d^3x,
      					\nonumber \\
    & &Q_{\rm B(KO)}' = \int K \stackrel{\leftrightarrow}{\del_0} B d^3x,
      					\nonumber \\
    & &Q_{\rm B(Y)~}' = \int c \stackrel{\leftrightarrow}{\del_0} Y_* d^3x.
\end{eqnarray}
The charge $Q_{\rm B(KO)}$ generates the BRST transformation for $A_\mu$,
$B$, $c$, $c_*$ and $\psi$ fields, that is, the same BRST transformation
as in the standard formalism. The transformation generated by $Q_{\rm
B(Y)}$ is the BRST transformation only for $Y$, $Y_*$, $K$ and $K_*$ fields.
Thus, $Q_{\rm B}$ in the last section can be decomposed as
\begin{equation}
                 Q_{\rm B} = Q_{\rm B(KO)} + Q_{\rm B(Y)}.
\end{equation}
The charge $Q_{\rm B(KO)}'$ generates the BRST transformation for $A_\mu$,
$B$ and $\psi$ but with $K$ and $K_*$ treated as their FP ghosts. Similarly,
$Q_{\rm B(Y)}'$ generates the BRST transformation for $Y$ and $Y_*$ with
$c$ and $c_*$ as their FP ghosts.

In last section, we have taken (\ref{QBphys}) as a physical subsidiary
condition. Instead of it, however, we may choose the condition as
\begin{eqnarray}
   & &Q_{\rm B(KO)} \left| {\rm phys} \right\rangle = 0, \nonumber \\
   & &Q_{\rm B(Y)}  \left| {\rm phys} \right\rangle = 0. \label{QBYphys}
\end{eqnarray}
The unphysical photons are removed by the first condition, while the gaugeon
modes by the second. Thus, this pair of conditions is much similar to
Yokoyama's pair (\ref{BRSphys}) and (\ref{Y*phys}) than the single condition
(\ref{QBphys}).
We denote the space of states satisfying (\ref{QBYphys}) by ${\cal V}_{\rm
phys}^{(\alpha)}$. As easily seen, this space is a subspace of ${\cal V}_
{\rm phys}$ defined in last section,
\begin{equation}
          {\cal V}_{\rm phys}^{(\alpha)}
                        \subset {\cal V}_{\rm phys}. \label{VsubV}
\end{equation}
We have used the index $(\alpha)$ in ${\cal V}_{\rm phys}^{(\alpha)}$ to
emphasize that the definition of the subspace ${\cal V}_{\rm phys}^{(\alpha)}$
depends on the gauge parameter $\alpha$. In fact, the BRST charges $Q_{\rm
B(KO)}$ and $Q_{\rm B(Y)}$ are not invariant under the q-number gauge
transformation (\ref{qgt2}). They transforms as
\begin{eqnarray}
   & &\hat Q_{\rm B(KO)} = Q_{\rm B(KO)} + \tau Q_{\rm B(KO)}',
   							\nonumber \\
   & &\hat Q_{\rm B(Y)} = Q_{\rm B(Y)} - \tau Q_{\rm B(KO)}',
   							\label{QBi_hat}
\end{eqnarray}
while their sum $Q_{\rm B}$ (and thus ${\cal V}_{\rm phys}$) remains
invariant.
In the following we show that the space ${\cal V}_{\rm phys}^{(\alpha)}$ has
the same structure with those of the physical subspace of the standard
formalism with the gauge parameter $a=\varepsilon\alpha^2$. In particular,
all of the Green functions for physical operators agree.

First we define a subspace ${\cal V}^{(\alpha)}$ of the total space
${\cal V}$ by
\begin{equation}
   {\cal V}^{(\alpha)}=\{ \left| \Phi \right\rangle\in {\cal V} ; ~
                Q_{\rm B(Y)} \left| \Phi \right\rangle =0 \}
                \subset {\cal V},
\end{equation}
which includes ${\cal V}_{\rm phys}^{(\alpha)}$ as a subspace since
${\cal V}_{\rm phys}^{(\alpha)}$ can be expressed as
\begin{equation}
   {\cal V}_{\rm phsy}^{(\alpha)}=
   \{ \left| \Phi \right\rangle\in {\cal V}^{(\alpha)} ; ~
                Q_{\rm B(KO)} \left| \Phi \right\rangle =0 \}
                \subset {\cal V}^{(\alpha)}
                             \label{QBKOphys}
\end{equation}
by definition. The space ${\cal V}^{(\alpha)}$ corresponds to the total
space of the standard formalism in the $a=\varepsilon\alpha^2$ gauge. And
thus, as seen from (\ref{QBKOphys}), ${\cal V}_{\rm phys}^{(\alpha)}$
corresponds to the physical subspace of the $a=\varepsilon\alpha^2$ standard
formalism. To see this we notice the following three facts:
\begin{enumerate}
  \item
   The equal-time commutation relations for the fields $A_\mu$, $B$, $c$,
   $c_*$ and $\psi$ are exactly the same with those of the standard formalism.
  \item
   If we take the matrix elements among the states of ${\cal V}^{(\alpha)}$,
   $A_\mu$, $B$, $c$, $c_*$ and $\psi$ satisfy the same field equations
   with those for the standard formalism in the $\varepsilon\alpha^2$-gauge.
   For example,
   \begin{equation}
        \left\langle \Phi_1 \right|
           \del^\mu A_\mu - \varepsilon\alpha^2 B
               \left| \Phi_2 \right\rangle = 0,
               \quad {\rm for~}
               \left| \Phi_1 \right\rangle,
               \left| \Phi_2 \right\rangle
               \in {\cal V}^{(\alpha)}
                                   \label{fact2}
   \end{equation}
   which is easily seen if we rewrite the equation (\ref{divA}) as
   \begin{equation}
      \del^\mu A_\mu - \varepsilon\alpha^2 B
              = i\varepsilon\alpha \{ Q_{\rm B(Y)}, K_* \}.
  \end{equation}
  \item
   Any state given by a product of the field operators $A_\mu$, $B$, $c$,
   $c_*$ and $\psi$ acting on the vacuum state is included in
   ${\cal V}^{(\alpha)}$, since these fields are $Q_{\rm B(Y)}$-singlets.
\end{enumerate}
The second fact can be easily understood if we express the Lagrangian
(\ref{LBRST}) as
\begin{equation}
  {\cal L} = {\cal L}_{\rm KO}^{(a=\epsilon\alpha^2)}
            + \left\{
              iQ_{\rm B(Y)}, \del_\mu K_* \del^\mu Y
              + {\varepsilon \over 2}K_*(Y_* + \alpha B)
              \right\},
\end{equation}
where ${\cal L}_{\rm KO}^{(a=\epsilon\alpha^2)}$ denotes the Lagrangian
of the standard formalism in $a=\varepsilon\alpha^2$ gauge.
The facts 1 and 2 mean that the field equations and the (four-dimensional)
commutation relations are the same with those of the standard formalism in
$\varepsilon\alpha^2$ gauge if their matrix elements are assumed to be taken
in ${\cal V}^{(\alpha)}$. Combining this with 3, we can conclude that
any vacuum expectation value of the products of $A_\mu$, $B$, $c$, $c_*$ and
$\psi$ fields coincides with that evaluated in the standard formalism.

It should be noted that the discussion in this section also applies to Izawa's
theory. Especially, his Lagrangian can be written as
\begin{equation}
  {\cal L}_{\rm Izawa} = {\cal L}_{\rm KO}^{(a=\alpha)}
            + \left\{
              iQ_{\rm B(Y)}, \del_\mu K_* \del^\mu Y + {1 \over 2}K_* B
              \right\},
\end{equation}
which leads to, for example, the equation similar to (\ref{fact2}).

\section{Path integral}

We start from the gauge invariant path integral
\begin{eqnarray}
    & &Z = \int {\cal D}A_\mu \exp \left[
                                     i \int {\cal L}_0 d^4x
                                   \right];
                                                   \label{Z0} \\
    & &{\cal L}_0 = - {1 \over 4} F^{\mu\nu}F_{\mu\nu},
\end{eqnarray}
where we have omitted the matter field $\psi$ for simplicity since its
presence is not essential in the following discussion. Before factoring out
the group volume by Faddev-Popov's trick \cite{FP}, we multiply $Z$ by unit
\begin{equation}
                1 = \det \Box^{-1} \cdot \det \Box ,
\end{equation}
and express the functional determinants as
\begin{eqnarray}
    & &\det \Box^{-1} =
          \int {\cal D}Y_* {\cal D}Y
               \exp \left[
                           i \int {\cal L}_{\rm Y} d^4x
                    \right],  \label{LY} \\
    & &\det \Box^{~~} =
          \int {\cal D}K_* {\cal D}K
               \exp \left[
                           i \int {\cal L}_{\rm K} d^4x
                    \right],
\end{eqnarray}
with
\begin{eqnarray}
    & &{\cal L}_{\rm Y} =
               \del_\mu Y_* \del^\mu Y + {\alpha_1 \over 2}Y_*^2 ,
                                            \label{LY*Y} \\
    & &{\cal L}_{\rm K} =
               \del_\mu K_* \del^\mu K ,
\end{eqnarray}
where $Y$ and $Y_*$ are bosonic scalar variables and $K$ and $K_*$ fermionic
scalar variables. In (\ref{LY}), $\alpha_1$ is a numerical parameter;%
\footnote{
        Although $\alpha_1$ is an arbitrary parameter, we can assume
        $\alpha_1=\varepsilon=\pm 1$ or $0$ without any loss of generality,
        since we may always use the field redefinitions
        $Y_* \rightarrow \pm e^\lambda Y_*$
        and
        $Y \rightarrow \pm e^{-\lambda} Y $
        which do not change the kinetic term in ${\cal L}_{\rm Y}$.
        }
a similar $K_*^2$-term cannot be introduced in ${\cal L}_{\rm K}$ because
of the anticommuting character of $K$ and $K_*$.
The gauge invariant path integral (\ref{Z0}) is now
\begin{equation}
  Z = \int {\cal D}A_\mu {\cal D}Y_* {\cal D}Y {\cal D}K_* {\cal D}K
    \exp \left[
         i \int ( {\cal L}_0 + {\cal L}_{\rm Y} + {\cal L}_{\rm K} )d^4x
         \right].
\end{equation}
We consider here the following gauge condition
\begin{equation}
   {\cal F} \equiv \del^\mu A_\mu - \alpha_2 Y_* = {\cal C}(x),
\end{equation}
where $\alpha_2$ is a numerical parameter and ${\cal C}(x)$ is an arbitrary
given function. Using Faddeev-Popov's trick for this gauge condition, we get
\begin{eqnarray}
  Z &=& \int {\cal D}A_\mu {\cal D}B {\cal D}Y_* {\cal D}Y
                   {\cal D}K_* {\cal D}K \Delta_{\rm FP} \nonumber \\
    & &\times \exp
                  \left[
                        i \int \{ {\cal L}_0 + {\cal L}_{\rm Y} +
                        {\cal L}_{\rm K} -B({\cal F} - {\cal C}) \} d^4x
                  \right],
                                     \label{Zb}
\end{eqnarray}
where we have expressed the $\delta$-functional $\delta[{\cal F}-{\cal C}]$
in the Fourier transformation form (integrated over $B$) and
$\Delta_{\rm FP}$ is the usual Faddeev-Popov determinant. Since the left
hand side is independent of the choice of the function ${\cal C}$, we can
take the 't~Hooft averaging \cite{tHooft} over ${\cal C}$. With the weight
functional
\begin{equation}
          \exp \left[
                      i \int {1 \over 2\alpha_3} {\cal C}(x)^2 d^4x
               \right],
\end{equation}
we finally get
\begin{equation}
     Z = \int {\cal D}A_\mu {\cal D}B {\cal D}c_* {\cal D}c
             {\cal D}Y_* {\cal D}Y {\cal D}K_* {\cal D}K 
         \exp\left[
                    i \int {\cal L}_{\rm YK} d^4x
            \right],
                                        \label{ZYK}
\end{equation}
where
\begin{eqnarray}
  {\cal L}_{\rm YK} &=& - {1\over 4}F_{\mu\nu}F^{\mu\nu}
                         + \del_\mu B A^\mu
                         + {\alpha_3 \over 2}B^2
                         + \alpha_2 BY_*
                         + \del_\mu Y_* \del^\mu Y
                                                   \nonumber \\
                     & & + {\alpha_1 \over 2}Y_*^2
                         - i \del_\mu c_* \del^\mu c
                         - i \del_\mu K_* \del^\mu K .
                                                 \label{LYK}
\end{eqnarray}

Thus we get the Lagrangian
with three parameters. Up to the FP ghost terms it coincides with the
three-parameter Lagrangian discussed by Yokoyama and Kubo \cite{YK},
the most general Lagrangian that includes the gaugeon fields.
Note that the Lagrangian (\ref{LYK}) is invariant under the BRST transformation
(\ref{BRST}). By choosing special values for these parameters,
we get the Lagrangian of Type I if $ \alpha_1=\pm1=\varepsilon,~\alpha_2=
\varepsilon\alpha,~\alpha_3=\varepsilon\alpha^2$, and of Type II if
$ \alpha_1=0, ~\alpha_2=1/2, ~\alpha_3=\alpha.$  (For other choice of the
parameters, see Ref.\cite{YK}. )

We have thus obtained the BRST symmetric Lagrangians of Type I and of
Type II (Izawa's Lagrangian) by path integral. The application to the
curved space-time case is straightforward.
{}From this derivation we easily understand the necessity of the gaugeon FP
ghosts $K$ and $K_*$. Without these FP ghosts the path integral differs by
$\det \Box$, which cannot be ignored in the background gravitational field.

\section{Summary}

By introducing gaugeon FP ghosts $K$ and $K_*$, we have given a
BRST symmetric version of Type I gaugeon formalism for quantum
electrodynamics. We also pointed out that the similar theory by Izawa can
be regarded as a BRST symmetrized Type II theory. The BRST symmetry enables
us to improve Yokoyama's subsidiary conditions (\ref{BRSphys}) and
(\ref{Y*phys}). We have replaced them by the single Kugo-Ojima
type condition (\ref{QBphys}), so that the formalism becomes applicable
to the case of background gravitational field.

Unlike Yokoyama's partially BRST symmetric theory, our fully symmetric theory
has the
physical subspace  ${\cal V}_{\rm phys}$ invariant under the q-number
gauge transformation. We have seen in \S4 that this gauge invariant space
${\cal V}_{\rm phys}$ includes various gauge variant subspaces
${\cal V}_{\rm phys}^{(\alpha)}$, which can be identified with the physical
subspaces of the standard formalism in $a=\varepsilon\alpha^2$ (or $a=\alpha$)
gauge.

A path integral derivation of our Lagrangian and Izawa's one is presented in
\S5. The derivation shows that the gaugeon FP ghosts $K$ and $K_*$ are
certainly necessary for the gaugeon formalism, especially in the background
gravitational field.

In this paper we restrict ourselves to the Abelian gauge theory.
We can also present BRST symmetric gaugeon formalism for the Yang-Mills
theory, which will be reported in the forthcoming paper.

\bigskip
\centerline{\bf Acknowledgements}
We would like to thank Professor K. Ishida for helpful comments and
encouragement.


%
\end{document}